\documentclass[twocolumn,superscriptaddress,nofootinbib]{revtex4-2}
\pdfoutput=1

\usepackage[utf8]{inputenc}
\usepackage{comment}
\usepackage[english]{babel}
\usepackage{amsfonts, amsmath, amssymb, amsthm}
\usepackage[colorlinks=true,breaklinks=true,linkcolor=black,citecolor=green,urlcolor=magenta]{hyperref}
\usepackage{graphicx}
\usepackage[export]{adjustbox}

\theoremstyle{remark}

\usepackage[capitalise]{cleveref}
\usepackage{algorithmicx}
\usepackage{algorithm}
\usepackage[noend]{algpseudocode}
\usepackage[normalem]{ulem}
\usepackage{booktabs}
\usepackage[thinlines]{easytable}
\usepackage{subcaption}
\usepackage{array}
\usepackage[table]{xcolor}
\usepackage{tikz}
\usepackage{qcircuit}
\usepackage{multirow}
\usepackage[symbol]{footmisc}

\algnewcommand\algorithmicinput{\textbf{Input:}}
\algnewcommand\Input{\item[\algorithmicinput]}

\algnewcommand\algorithmicoutput{\textbf{Output:}}
\algnewcommand\Output{\item[\algorithmicoutput]}

\usepackage{bm}
\usepackage{braket}

\newcommand{\nocontentsline}[3]{}
\newcommand{\tocless}[2]{\bgroup\let\addcontentsline=\nocontentsline#1{#2}\egroup}

\newcommand{\timesdiv}{{\;\mathbin{\vcenter{\hbox{%
   $\begin{array}{@{}c@{}}{\times}\\[-1ex]{\div}\end{array}$}}}\;}}

\begin{document}

\title{Pauli Network Circuit Synthesis with Reinforcement Learning}

\def\equalauth{\dag}

\author{Ayushi Dubal\textsuperscript{\equalauth}}
\affiliation{IBM Quantum, IBM T.J. Watson Research Center, Yorktown Heights, NY 10598}
\affiliation{Department of Computer Sciences, University of Wisconsin-Madison, Madison, WI 53706}

\author{David Kremer\textsuperscript{\equalauth}}
\email{david.kremer@ibm.com}
\affiliation{IBM Quantum, IBM T.J. Watson Research Center, Yorktown Heights, NY 10598}

\author{Simon Martiel}
\affiliation{IBM Quantum, IBM France Lab, Saclay, France}

\author{Victor Villar}
\affiliation{IBM Quantum, IBM T.J. Watson Research Center, Yorktown Heights, NY 10598}

\author{Derek Wang}
\affiliation{IBM Quantum, IBM T.J. Watson Research Center, Yorktown Heights, NY 10598}

\author{Juan Cruz-Benito}
\affiliation{IBM Quantum, IBM T.J. Watson Research Center, Yorktown Heights, NY 10598}

\footnotetext[2]{These authors contributed equally to this work.}

\begin{abstract}

We introduce a Reinforcement Learning (RL)-based method for re-synthesis of quantum circuits containing arbitrary Pauli rotations alongside Clifford operations. By collapsing each sub-block to a compact representation and then synthesizing it step-by-step through a learned heuristic, we obtain circuits that are both shorter and compliant with hardware connectivity constraints. We find that the method is fast enough and good enough to work as an optimization procedure: in direct comparisons on 6-qubit random Pauli Networks against state-of-the-art heuristic methods, our RL approach yields over 2× reduction in two-qubit gate count, while executing in under 10 milliseconds per circuit. We further integrate the method into a collect-and-re-synthesize pipeline, applied as a Qiskit transpiler pass, where we observe average improvements of 20\% in two-qubit gate count and depth—reaching up to 60\% for many instances—across the Benchpress benchmark. These results highlight the potential of RL-driven synthesis to significantly improve circuit quality in realistic, large-scale quantum transpilation workloads.

\end{abstract}


\maketitle

\tocless\section{Introduction}\label{section:introduction}

Efficient transpiling of quantum circuits is required to bridge the gap between abstract quantum algorithms and practical implementation. Quantum circuit optimization is a central topic in quantum computing, especially for current quantum devices that suffer from relatively high gate error rates and limited qubit connectivity. The task of reducing the number of gates, especially expensive multi-qubit gates, and shortening circuit depth is crucial for mitigating the accumulation of errors, thus improving the fidelity of quantum algorithms. Modern quantum compilers, or “transpilers,” such as Qiskit \cite{javadi2024quantum} employ a wide range of techniques to systematically reduce circuit size and adhere to hardware constraints such as limited qubit connectivity. 

Achieving global transpiling optimality for large circuits is widely recognized to be infeasible due to the exponential growth of the search space \cite{ge2024quantum,karuppasamy2025comprehensive}. Consequently, local optimizations, where a circuit is decomposed into smaller blocks and each is optimized individually, have emerged as a practical strategy \cite{weiden2022widequantumcircuitoptimization, Sivarajah_2021}. 

A common way to perform the optimization of the sub-circuits in each block is to just collapse the circuit into some representation (e.g. unitary matrix, Clifford tableau, etc...) and then perform re-synthesis of the operation \cite{Clark_2024, liu2020relaxedpeepholeoptimizationnovel}. This can be done at different stages of the transpilation. One could, for instance, do this before the routing stage, obtaining better opportunities for optimization on the “pure” circuit but risking larger overhead after routing. One could also do this after the routing stage, where the circuit is already mapped to hardware connectivity and “scrambled”, but where any improvements directly translate to improvements in the final circuit \cite{li2022paulihedral, Jin:2023mil}. In this work, we focus on the post-routing scenario.

This post-routing re-synthesis approach has the advantage of potentially reaching optimal implementations for each sub-circuit, but for practical application to large circuits it requires the re-synthesis method to fulfill the following:
\begin{itemize}
    \item Implement the exact unitary of the original sub-circuit so that it can replace it.
    \item Preserve the coupling map, so that there is no further need of routing.
    \item Be “optimal enough” so that for each re-synthesized block there is a reasonable chance that it will improve the original sub-circuit.
    \item Be “fast enough” that it can be ran for many blocks in the circuit.
\end{itemize}

In earlier work \cite{kremer2024practical}, it was shown how these requirements can be satisfied with a Reinforcement Learning (RL)-based approach for Clifford blocks up to 10 qubits, enabling the local optimization of larger circuits through sub-block re-synthesis. However, this earlier method restricted attention to purely Clifford blocks, omitting possible gains from incorporating more general operations. 

Here, we extend this result to the re-synthesis of Pauli Networks up to 6 qubits, allowing the collected blocks to contain a combination of Clifford operations and Pauli rotations. This enables the optimization of arbitrary (unitary) blocks, potentially containing unbound parametric rotations, and allows peephole optimizations that are not possible when only collecting Clifford blocks. As described in \cite{kremer2024practical}, the RL method performs the synthesis step-by-step by selecting the most adequate gate at each step through a learned heuristic. We find that, as in the case of purely Clifford blocks, it is possible for the RL model to learn a good heuristic for Pauli Network blocks that achieves short circuits with very high success rate while also following a coupling map.

There is extensive work on heuristic algorithms for Clifford circuit synthesis with and without connectivity restrictions \cite{bravyi2021clifford, debrugière2022graphstate}, optimal databases of Clifford circuits \cite{Bravyi_2022}, and methods for encoding Clifford synthesis as a SAT problem \cite{peham2023depthoptimalsynthesiscliffordcircuits, Schneider_2023}. However, these methods are limited to Clifford circuits, and cannot perform optimization accross non-Clifford gates. For synthesis of Pauli rotations, Rustiq \cite{debrugière2024faster} provides strong results for two-qubit gate count and depth synthesis for full connectivity, and potentially scales to hundreds of qubits, but it does not accommodate for connectivity restrictions and does not efficiently implement the overall Clifford operator. Other methods such as \cite{Mukhopadhyay_2023} perform synthesis of sequences of rotations, but focus on the fault tolerant case and optimize the number of non-Clifford gates. Although the re-synthesis of full unitary circuits is possible with methods such as approximate compiling \cite{madden2021bestapproximatequantumcompiling}, these methods scale poorly, do not implement the exact unitary (even if they may get arbitrarily close) and cannot handle parametric rotations.
Our approach marks the first non-approximate peephole optimization approach applicable to a universal class of circuits (Clifford + single-qubit rotations), including parametrized circuits containing unbound parameters.

We validate our method’s performance in two ways. First, we benchmark the RL-based synthesis against state-of-the-art heuristics (including Qiskit \cite{javadi2024quantum} and Rustiq \cite{debrugière2024faster}) on up to 6-qubit random Pauli Networks. We observe meaningful reductions of 50\% on average in gate count and circuit depth once routing overheads are taken into account. Second, we incorporate the approach into a collect-and-re-synthesize procedure for full-circuit optimization, and run it as a post-transpilation optimization pass on the Benchpress benchmark \cite{nation2024benchmarking}. Our results show an average improvement of 20\% in two-qubit gate count and depth on average across the benchmark, reaching up to 40\%-60\% for a meaningful number of cases.

The remainder of this paper is organized as follows. In Methods, we detail how we adapted the RL synthesis procedure from \cite{kremer2024practical} to handle Pauli Network blocks, and describe the overall collect-and-re-synthesis pipeline as a Qiskit-compatible transpiler pass. In Results, we present the outcomes of two primary benchmarks: one focusing on direct synthesis of random 6-qubit Pauli Networks, and another illustrating the end-to-end improvement in the Benchpress benchmark. Finally, in Conclusion, we discuss our findings and potential future avenues, such as scaling to larger sub-circuits and exploring different collection strategies.

\vspace{2em}
\tocless\section{METHODS}\label{section:methods}

\subsection{RL-based circuit synthesis}

In this work, we follow the RL synthesis method described in \cite{kremer2024practical}, with some modifications to accommodate the properties of the problem. 

At the core of the algorithm, we define a neural network that takes a representation of the state (here, a Pauli Network) and generates a distribution over all possible gate choices. The set of possible gates can be anything as long as it can generate the full Clifford group for a given number of qubits. In particular, we can restrict the placement of the two-qubit gates to the pairs of qubits allowed by a given coupling map.

\begin{figure}
\begin{subfigure}[b]{0.45\textwidth}
    \centering
    \includegraphics[width=\linewidth]{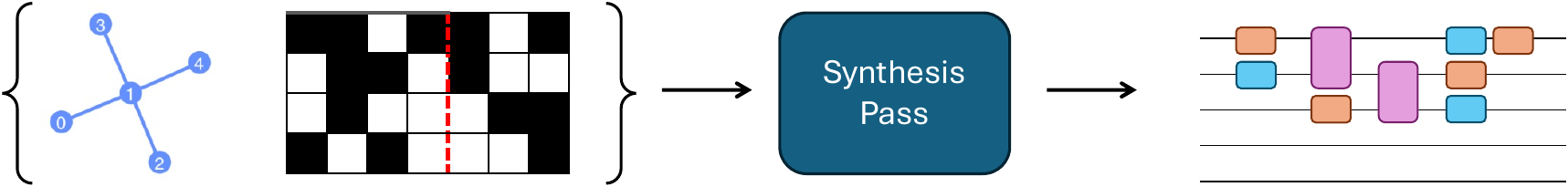}
    \caption{Pauli Network synthesis at a high level, where a coupling map and a Pauli Netowork are provided as inputs, and the RL-based procedure generates a circuit that implements the Pauli Network following the coupling map.}
    \label{fig:workflow}
\end{subfigure}
\begin{subfigure}[b]{0.45\textwidth}
    \centering
    \includegraphics[width=\linewidth]{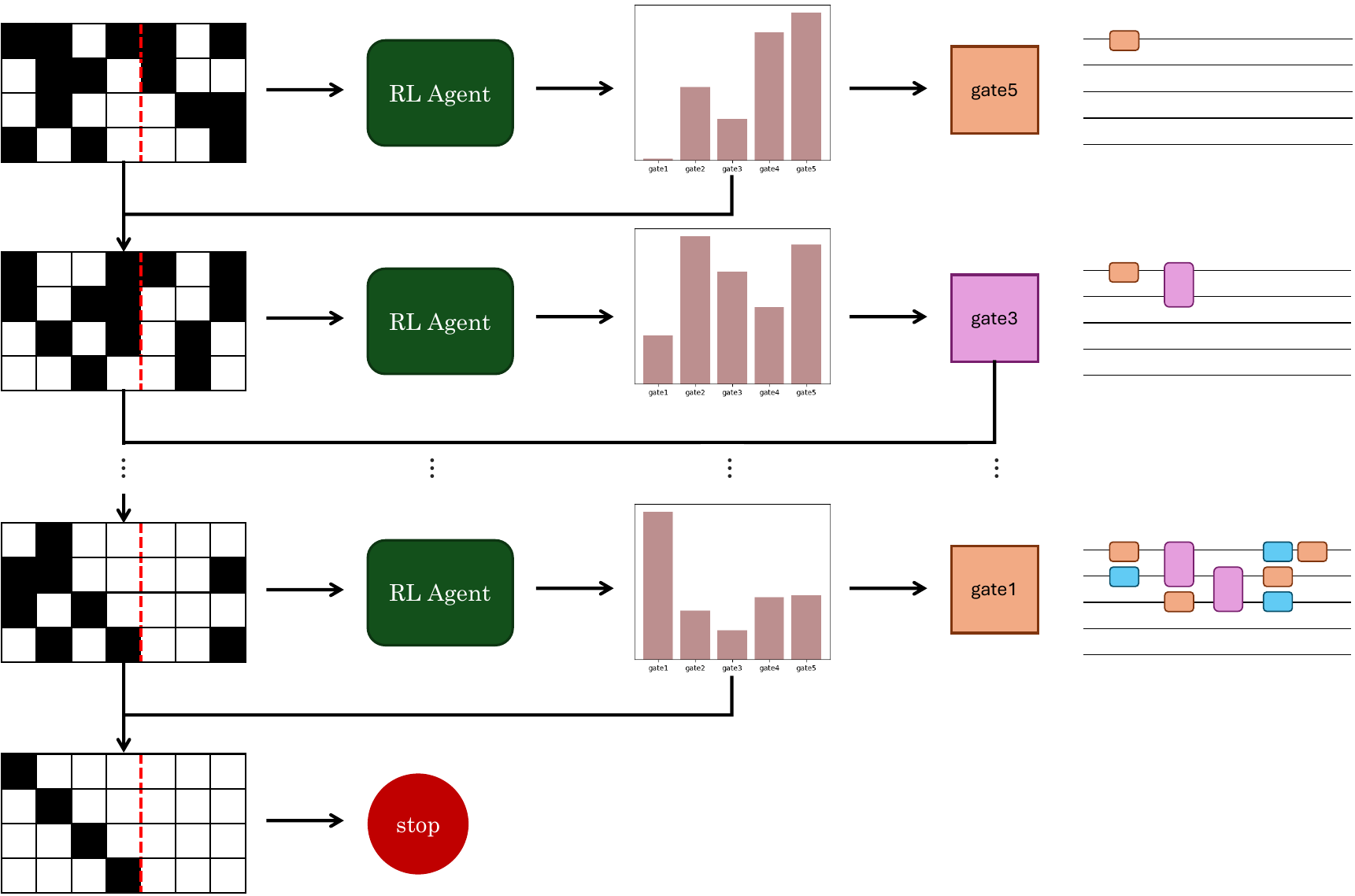}
    \caption{RL-based synthesis procedure. See main text for details.}
    \label{fig:inference}
\end{subfigure}
\caption{}
\end{figure}

When given an initial state, this network is used to generate a sequence of gates that implement the given Pauli Network by iteratively generating gates: 
\begin{enumerate}
    \item The current state is used as input to the network, which provides probabilities for each possible gate.
    \item A gate is chosen by sampling from the distribution.
    \item The gate is used to update the current state to generate the next state in the sequence.
    \item The process is repeated until the terminal state is reached (here, the identity Pauli Network).
\end{enumerate}
This procedure is depicted in Figure \ref{fig:workflow}. 

In order to train this network, we follow standard RL practices and train the model with Proximal Policy Optimization (PPO) \cite{schulman2017proximal}. We first start with a ``blank" network, and collect synthesis training data by running the synthesis on randomly generated Pauli Networks. For each synthesis run, a reward is provided based on the success of the process and the quality of the circuit generated. The network is then updated through gradient descent to maximize this reward function. This collect and train procedure is repeated until the network reaches the desired synthesis quality.

In the following subsections we provide more details on the state representation and the training procedures we follow.

\subsection{State representation}

We briefly recall the Pauli Network framework and the natural extension of the Clifford tableaux framework to this more general set of operators.

\textbf{Pauli Networks.} Any quantum circuit composed of Clifford gates and (single qubit) Pauli rotations of arbitrary angles can be normalized into a sequence of Pauli rotations followed by a Clifford operator. This straightforward normalization process is commonly used in circuit synthesis literature to deal with arbitrary Clifford + RZ/T circuits \cite{Martiel2022architectureaware, Litinski2019gameofsurfacecodes} and can be efficiently computed by iterating through the circuit, using Clifford gates to update a running Clifford tableau (which takes $O(n)$ operations) and conjugating each (single qubit) Pauli rotation through the running tableau (which takes $O(n)$ operations for single qubit gates and $O(n^2)$ operations for arbitrary Pauli rotations). Overall, this normalization process costs $O(nm)$ operations where $n$ is the number of qubits and $m$ the number of gates in the case of single qubit Pauli rotations and $O(n^2m)$ the general case. After this process, our input circuit $U$ can be written as:
\begin{align}
    U &= C \prod_{i=1}^m R_{P_i}(\theta_i) \label{eq:pauli_network_form}
\end{align}
where $C$ is a Clifford operator, $P_i$ are Pauli operators in $\{I, X, Y, Z\}^{\otimes n}$, and $\theta_i$ are real numbers (or parameters in the case of unbounded parametrized circuits).
Notice also that in the sequence $\prod_{i=1}^m R_{P_i}(\theta_i)$, rotations might commute, implying that the order in the product can be relaxed \emph{up to commutation relations} between rotations. In other words, one can implement this product of rotations in any order $\prec$ that satisfies $\forall 1 \leq i < j \leq m$, $[P_i, P_j] \neq 0 \implies i \prec j$. This constraint can be easily captured by a Directed Acyclic Graph (DAG) $D=(\{1, ..., m\}, \{(j, i), \forall 1 \leq i < j \leq m, [P_i, P_j] \neq 0\})$, or less formally, by a graph whose vertices are rotation indices and where a $j$ links to $i$ if and only if $i$ comes before $j$ and $P_i$ and $P_j$ anti-commute.

\begin{figure}[h!]
    \centering
    \begin{subfigure}[b]{0.35\textwidth}
        \includegraphics[width=\linewidth]{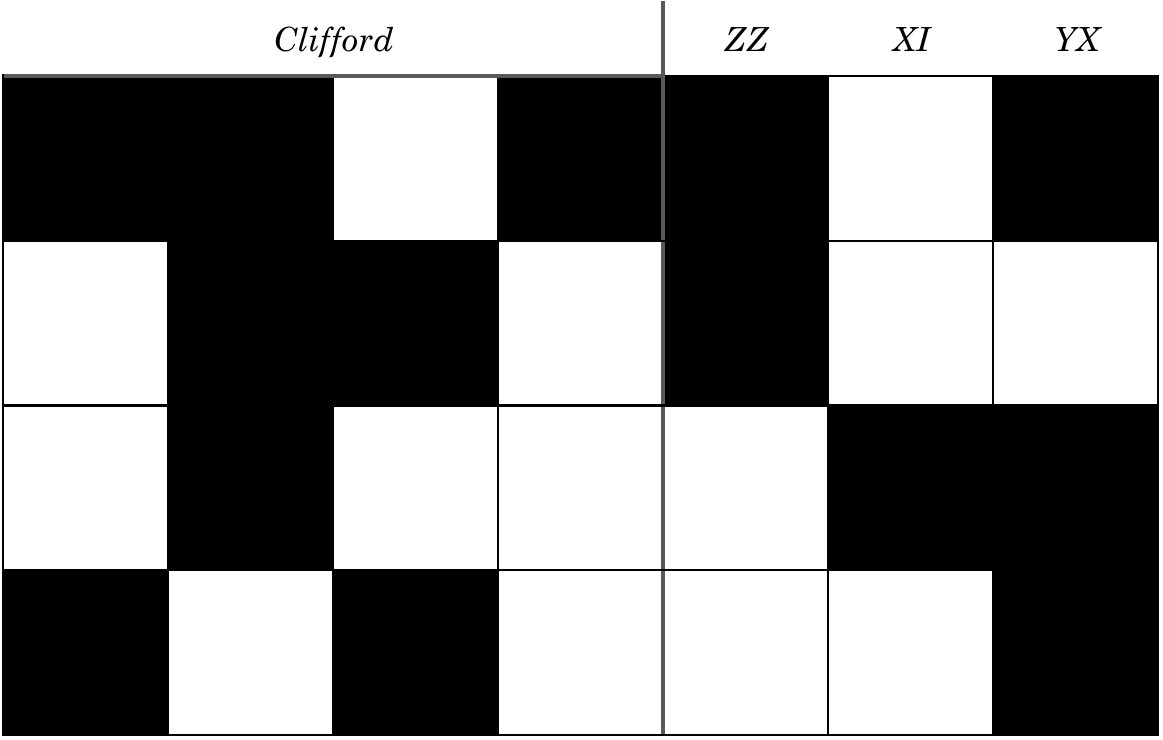}
        \caption{Tableau representation of Clifford followed by $p=3$ Pauli strings acting on $n=2$ qubits.}
        \label{fig:pauli_tableau}
    \end{subfigure}
    \hfill
    \begin{subfigure}[b]{0.35\textwidth}
        \includegraphics[width=\linewidth]{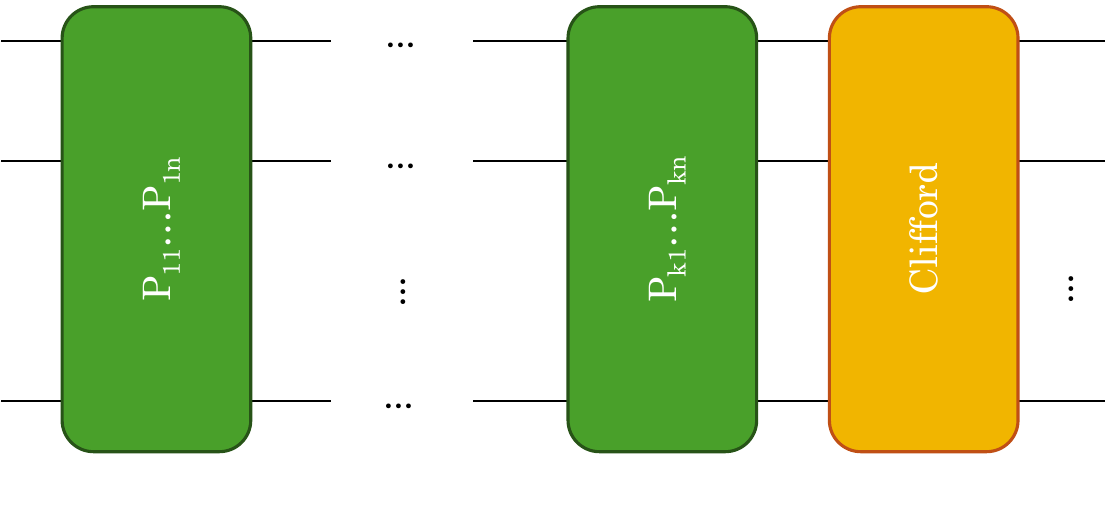}
        \caption{Structure of Pauli Network block}
        \label{fig:pn_structure}
    \end{subfigure}
    \caption{}
\end{figure}

\textbf{Representation.} Any operator described by Eq. \ref{eq:pauli_network_form} can be described using $2n + m$ Pauli operators: $2n$ operators to describe $C$ and $m$ operators to describe the $P_i$s. 
A Pauli operator acting on $n$ qubits, which has the form $P_1\otimes P_2\otimes\cdots \otimes P_n$, where $P \in \{I, X, Y, Z\}$ can be vectorized as follows. Each Pauli $P_i$ is encoded as a 2-bit string $z_ix_i$, with $z_i$ representing the $Z$-component and $x_i$ representing the $X$-component. Thus, the mapping is $I \rightarrow 00, X \rightarrow 01, Y \rightarrow 11, Z \rightarrow 10$. The Pauli operator $P_1P_2…P_n$ is written as the vector $z_1 z_2 .. z_n | x_1 x_2 … x_n$. 
Overall, we thus store a $2n \times (2n + m)$ bit table storing the Clifford tableau representing $C$ in its first $2n$ columns and the $P_i$ operators in the last $m$ columns.
This tableau can be visualized in \ref{fig:pauli_tableau}. We also extend the representation with the DAG described in the previous section.

\begin{figure}[ht!]
    \centering
    \input{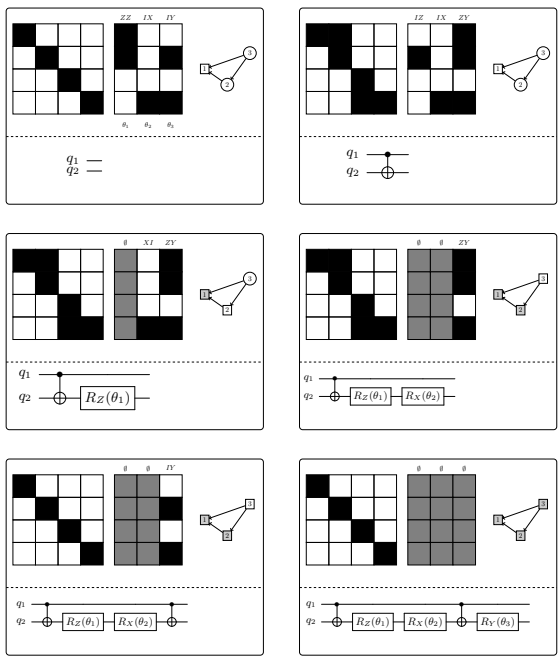}
    \caption{Example of agent update. The agent starts by adding a CNOT to the output circuit. The tableau is updated accordingly. We then notice that the first rotation (which lies in the front layer of the DAG) became trivial (i.e. a single qubit rotation). The corresponding rotation is added to the output circuit, the Pauli is cleared, and $1$ is removed from the DAG. The second rotation is now part of the front layer and is also trivial, leading to the same treatment, finalizing the update. The agent inserts a last CNOT, the tableau is updated. The third rotation is now part of the front layer and trivial, leading to the injection of a last $R_Y$ into the output circuit. After those two moves, the Clifford tableau (left tableau) is trivial (i.e. implements the identity) and the DAG is empty: the synthesis process is over.}
    \label{fig:agent_update}
\end{figure}

\textbf{Representation update and agent actions.}
This tableau is evolved at every step by the RL agent. The RL agent can decide to insert a Clifford gate in the output circuit. The $2n + m$ Pauli operators are updated by conjugation by this Clifford gate. Whenever a $P_i$ operator satisfies the two following conditions:
\begin{enumerate}
    \item $i$ has not outgoing edges in the anti-commutation DAG (i.e. $i$ is part of the front layer of the DAG)
    \item $P_i$ has weight exactly $1$
\end{enumerate}
column $P_i$ is set to zero, the corresponding single qubit rotation is added to the output circuit, and $i$ is removed from the DAG. This operation is repeated until no $P_i$ can satisfy the two conditions. An example of the state updates is depicted in Figure \ref{fig:agent_update}

\subsection{Training}

\textbf{Reward function.} The reward function is straightforward - there is a large positive reward for the case that the target operator is synthesized. For every gate added to the circuit, there is a small negative reward. There is a slightly larger penalty for two qubit gates. 

\textbf{Training data generation.} During training, batches of operators to be synthesized are randomly generated. The initial state of the system is determined by the generated Clifford and Pauli operator list. The target generation is done as per a parameter called ‘difficulty’ \cite{10.5555/3455716.3455897}. At difficulty $d$, we generate a sequence of $d$ random Pauli operators as the target for the agent to learn. 

We define two dimensions of difficulty to the synthesis problem - the difficulty of the Clifford and the difficulty of the Pauli operators. For the Clifford, we start with an identity tableau, and apply $d_C$ operations drawn randomly from the possible set of gates; this ensures that the Clifford part is implementable in at most $d_C$ steps. For the Pauli part, we start with an empty list of Pauli strings, and add a random Pauli string for each $d_P$. In order to keep the training progressing, we start from $d_C = 1$ and progressively increase it by 1 when the model reaches a given success rate threshold. The value of $d_P$ is set by scaling $d_C$ with a coefficient $\delta$ - 
\begin{equation}
    d_P = \left\lfloor\frac{d_C}{\delta}\right\rfloor
\end{equation}. For the results presented in this paper, we use $\delta = 8$.

\textbf{Network architecture.} In order to encode it as an input to the neural network, the Pauli Network is converted to a boolean matrix of shape $2N$ by $2N + P$ where $N$ is the number of qubits and $P$ the number of Pauli strings. For the Pauli terms that commute, their ordering in the matrix is determined by the order in which they were collected/generated, but we observed that randomly changing this ordering does not affect the results even after training.

The network is composed of 3 layers: a first Conv1d layer that applies an 1d convolution across the second dimension and flattens the output, followed by two fully connected layers that at the end output a logit for each possible gate. In order to handle arbitrarily long sequences of Pauli strings, we set a horizon $H$ and cut the matrix to a $2N$ by $2N + H$ shape picking the first $H$ Pauli strings.

\subsection{Circuit optimization through collection and re-synthesis}

We use the models trained to do Pauli Network circuit synthesis up to a given (small) number of qubits as the core component of an optimization procedure that can be applied to circuits of arbitrary size. The procedure is intended to oprimize the input circuit after it has been mapped to a coupling map.

The most basic version of this procedure has three steps:
\begin{enumerate}
    \item \textbf{Collection.} A list of sub-circuits is selected from within the circuit, of sizes up to the largest number of qubits that can be synthesized by the RL algorithm. These sub-circuits are converted into a Pauli Network representation.
    \item \textbf{Model selection.} An RL model is selected for each of the sub-circuits, depending on the number of qubits and the underlying graph on which the sub-circuit falls.
    \item \textbf{Re-synthesis.} Each sub-circuit is re-synthesized with the RL method. If the new sub-circuit improves the original one by a metric of choice (e.g. two-qubit depth or count) then it is replaced. 
\end{enumerate}

Note that while step 3 guarantees that the overall circuit will at worst stay the same in terms of two-qubit gate count, it is still possible that global depth is increased (even when all replacements decrease or maintain depth), although we have found this not to be the case for most circuits we tested.

\begin{figure}[ht]
    \centering
    \includegraphics[width=1\linewidth]{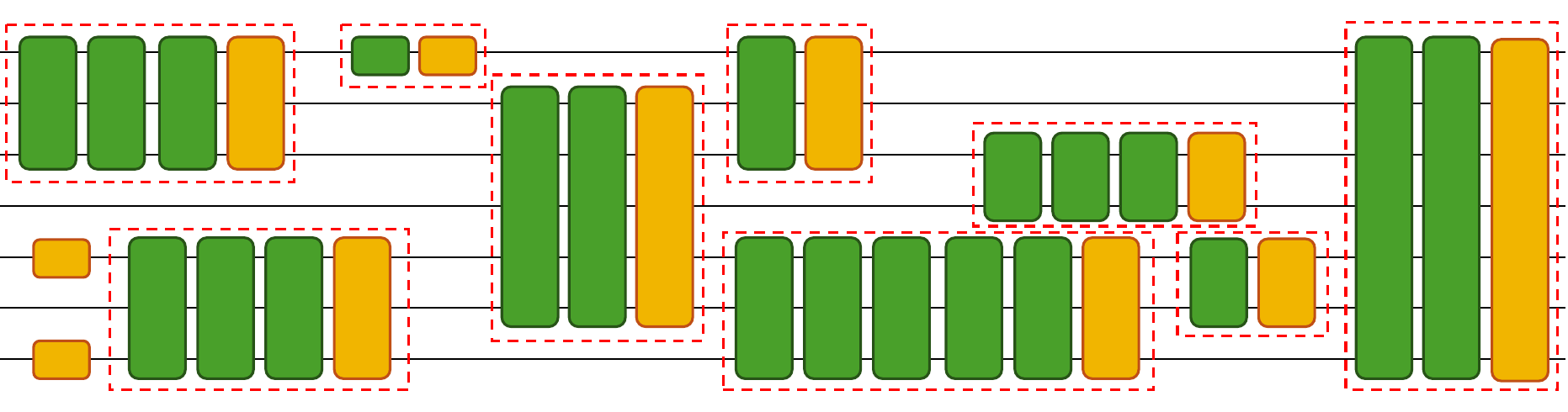}
    \caption{Pauli Network collection pass on an arbitrary circuit. The collected sub-circuits are to be re-synthesized by the RL agent to optimize on count or depth.}
    \label{fig:optimization_pass}
\end{figure}

For the collection procedure, we use a greedy method that starts a block from a ``seed gate" at the beginning of the circuit, and then starts adding gates to the block following the circuit's DAG using a depth first search, until the maximum number of sub-circuit qubits is reached. It then repeats by taking the next gate that is not part of a block, and generates blocks until all gates are part of some block. We also use a variant of this, but where the collection is done starting from the end of the circuit (left to right) instead of from the front (right to left). Figure \ref{fig:optimization_pass} shows a depiction of this procedure.

\tocless\section{RESULTS}\label{section:results}

\subsection{Results for synthesis}

We train models for all subgraphs of the heavy-hex coupling map for 4, 5 and 6 qubits (see \cite{kremer2024practical} for the list of subgraphs). We run a ``native synthesis" benchmark where we first generate a random Pauli Network of a given size, and then run a list of algorithms that do the synthesis into a chosen coupling map. The algorithms must reproduce the full operation completely, and must follow the coupling map. We then measure the total synthesis time, number of two-qubit gates, and number of two-qubit layers.

For the benchmark, we compare the following algorithms:
\begin{enumerate}
    \item \texttt{QiskitSDK Transpiler (Default, lvl2)} - The Pauli list component is synthesized following \cite{li2022paulihedral}, and the Clifford component with the Greedy method \cite{bravyi2021clifford}. The output is then routed to the coupling map with Sabre \cite{li2019tackling,zou2024lightsabre}.
    \item \texttt{QiskitSDK Transpiler (Rustiq synthesis, lvl2)} - The same as the previous one, but where the Pauli list is synthesized with Rustiq \cite{debrugière2024faster} instead.
    \item \texttt{RL\_n} - The Pauli Network is synthesized by performing `n' runs of the RL algorithm and taking the best result.
\end{enumerate}

Note that after transpiling the first two methods do not fully reproduce the original operator, only up to a final permutation or final Clifford operator (in the case of \cite{debrugière2024faster}). If we were to use them as part of a collect and re-synthesis procedure, we would have to append this extra permutation/Clifford, further increasing the circuit size.

For the generation of the target Pauli Networks we combine the following parameters:

\begin{enumerate}
\item \texttt{num\_rotations} - The number of Pauli strings.
\item \texttt{input\_paulis} - Whether the randomly generated Pauli strings are from the full set \texttt{\{I,X,Y,Z\}} or only \texttt{\{I,Z\}}.
\item \texttt{coupling\_map} - The target coupling map to synthesize to.
\end{enumerate}

\begin{figure*}[t]
    \centering
    \begin{minipage}{0.48\textwidth}
        \centering
        \includegraphics[width=1.0\linewidth]{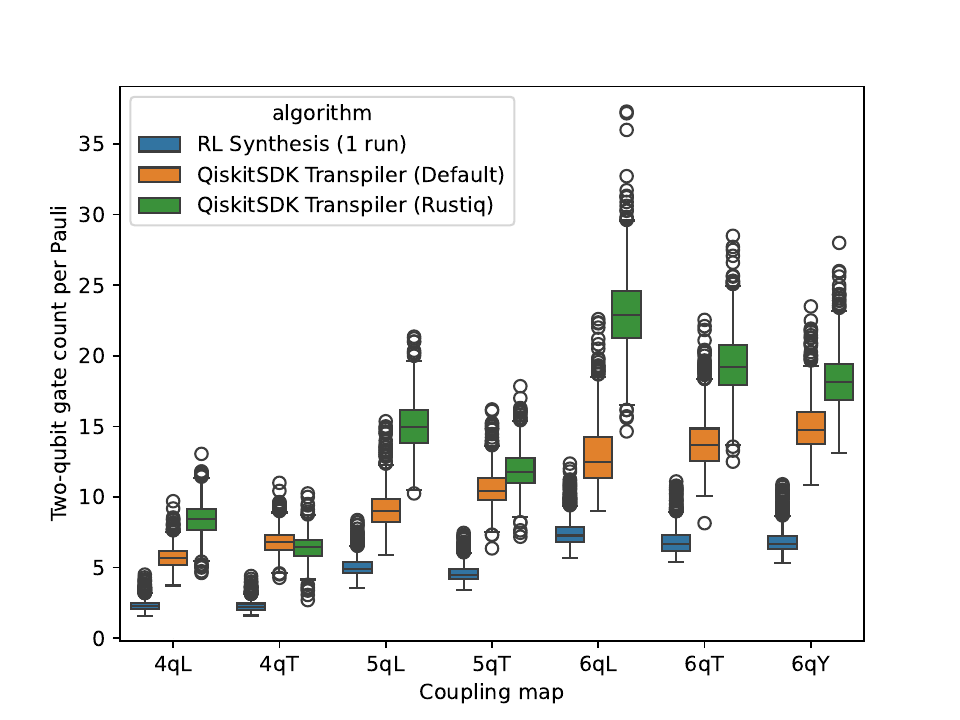}
        \caption{Number of two-qubit gates per Pauli rotation for the different algorithms and coupling maps, for full IXYZ Paulis.}
        \label{fig:count2q_boxplot}
    \end{minipage}\hfill
    \begin{minipage}{0.48\textwidth}
        \centering
        \includegraphics[width=1.0\linewidth]{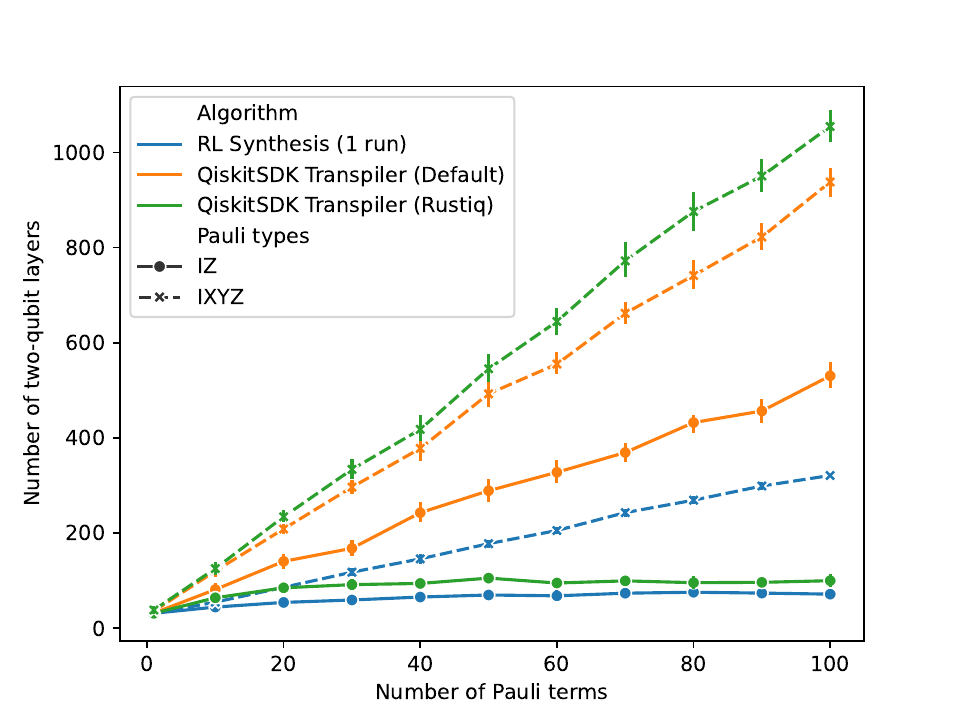}
        \caption{Number of two-qubit layers achieved by the different algorithms vs the number of Pauli strings for a ``5qT" coupling map. }
        \label{fig:count2q_5qT}
    \end{minipage}
\end{figure*}

In Figure \ref{fig:count2q_boxplot} we show the average number of two-qubit gates achieved by the different algorithms for the different coupling maps, normalized by the number of Pauli strings. The RL synthesis algorithm generates circuits that in general contain 50\% less two-qubit gates than the second best algorithm. We also observe that, although Rustiq generates very small circuits, the subsequent routing makes the final circuit larger than with the default Qiskit transpiling pipeline, likely due to the default Qiskit synthesis being easier to route to coupling maps that are close to linear connectivity.

In Figure \ref{fig:count2q_5qT} we show how the number of two-qubit layers evolves with the number of Pauli terms for the different algorithms, in the cases where the Paulis only have IZ terms and where they have all IXYZ terms, for a ``5qT" coupling map. For the IXYZ benchmark, we see that the depth scales linearly with the number of Paulis for all algorithms; Qiskit perfomrs similarly with default and Rustiq synthesis, and RL synthesis generates circuits around 4x shallower. For the IZ benchmark, however, Qiskit with default synthesis scales linearly, but Rustiq and the RL quickly reach a plateau after around 20 Paulis. 

Regarding synthesis time, we found that all three methods performed within the same order of magnitude, averaging around 0.5 to 3 milliseconds per Pauli. The full results for time, two-qubit count and depth can be found in the Appendix.

\subsection{Results for large circuit optimization}

We define the following building blocks that we use for optimization:
\begin{itemize}
    \item \texttt{collect\_{forward/backward}}: Parses through the circuit, collecting Pauli Network blocks starting from left to right (forward), or right to left (backward).
    \item \texttt{synthesize\_{runs}}: Performs RL synthesis of each of the collected blocks for ``runs" runs per block, selecting the best sub-circuit in each case, and replacing the original one if it is improved.
\end{itemize}

Based on this, we define four different optimization workflows with increasing computational cost and optimization performance:
\begin{itemize}
    \item \texttt{rl\_4q0\_10}: [\texttt{collect\_forward}, \texttt{synthesize\_10}]
    \item \texttt{rl\_4q1\_100}: [\texttt{collect\_forward}, \texttt{synthesize\_100}, \texttt{collect\_backward}, \texttt{synthesize\_100}]
    \item \texttt{rl\_4q3\_100}: [\texttt{rl\_4q1\_100}, \texttt{rl\_4q1\_100}, \texttt{rl\_4q1\_100}]
    \item \texttt{rl\_all\_100}: [\texttt{rl\_4q3\_100}, \texttt{rl\_5q1\_100}, \texttt{rl\_6q1\_100}]
\end{itemize}
where the procedures are applied sequentially to an input circuit. Note that here the \texttt{rl\_4q*} use the 4-qubit models, and \texttt{rl\_5q1\_100} and \texttt{rl\_6q1\_100} use the 5- and 6-qubit models respectively. We alternate forward and backward collection in order to increase the chance of collecting different blocks at each iteration and increase the potential for optimization. We find that repeatedly applying this forward and backward collection and re-synthesis often yields increasingly better results at the expense of more optimization time.

We run these optimization methods on the Benchpress benchmark \cite{nation2024benchmarking}. For measuring the method's potential for circuit optimization post-transpilation, we first run the QiskitSDK transpiler (optimization level 2, software version 1.3) and then run each of the optimization methods independently on the transpiled circuit. 

In Table \ref{tab:benchpress_results} we show the reduction factors for two-qubit count and two-qubit depth obtained on average for each of the subsets of the Benchpress benchmark. We see that overall there is a ~10\% to ~20\% improvement in two-qubit gate counts for the lightest optimization, and ~15\% to ~30\% for the heaviest one. The full results for each of the circuits can be found in the Appendix.

\begin{figure}[tp]
    \centering
    \includegraphics[width=\linewidth]{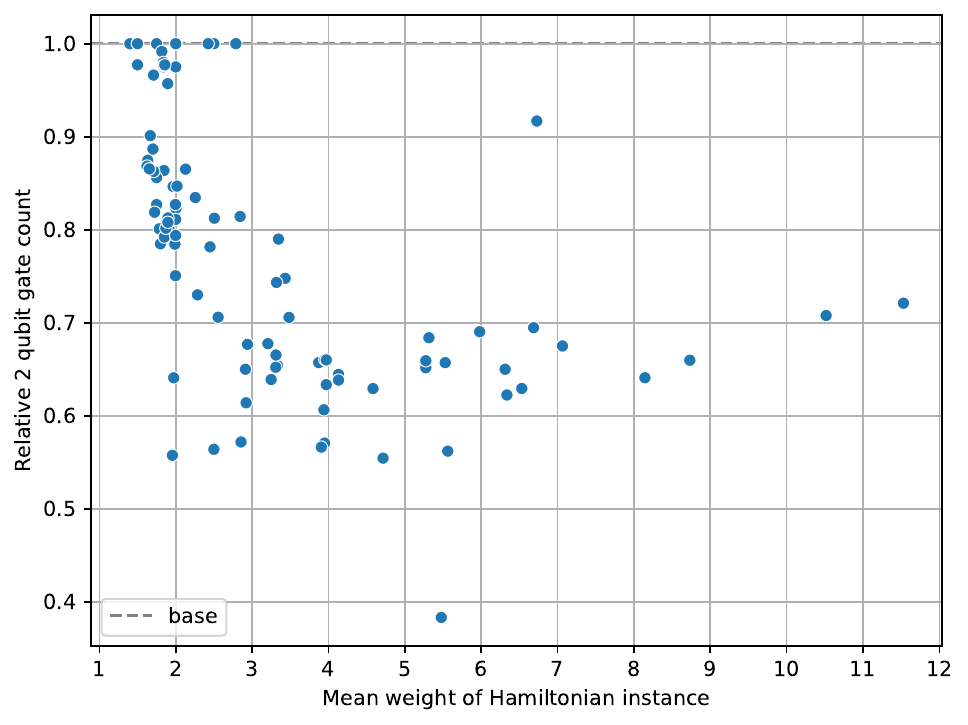}
    \caption{A comparison of the 2 qubit gate count improvement of the \texttt{rl\_all\_100} workflow over Qiskit level 2 against the mean hamming weight of the input Paulis. We see that as the mean hamming weight of the input Pauli list increases, there is a greater improvement, which eventually plateaus out at about $40\%$.}
    \label{fig:ham_wt_vs_count}
\end{figure}

We also note that there is a large variance in the results, where some of the circuits see no improvement at all, while the ones that improve often do so by a large amount (around ~30\% to ~50\%). To get a closer look into what factors influence this, we plot the improvement obtained against the mean Pauli weight of the circuit for the HamLib subset of Benchpress (the circuits here are originally defined as a list of Pauli strings), as shown in Figure \ref{fig:ham_wt_vs_count}. We see that that most of the low-improvement circuits are below a Pauli weight of 3 and almost all the circuits above that show improvements beyond 30\%.

\begin{figure}[tp]
    \centering
    \includegraphics[width=\linewidth]{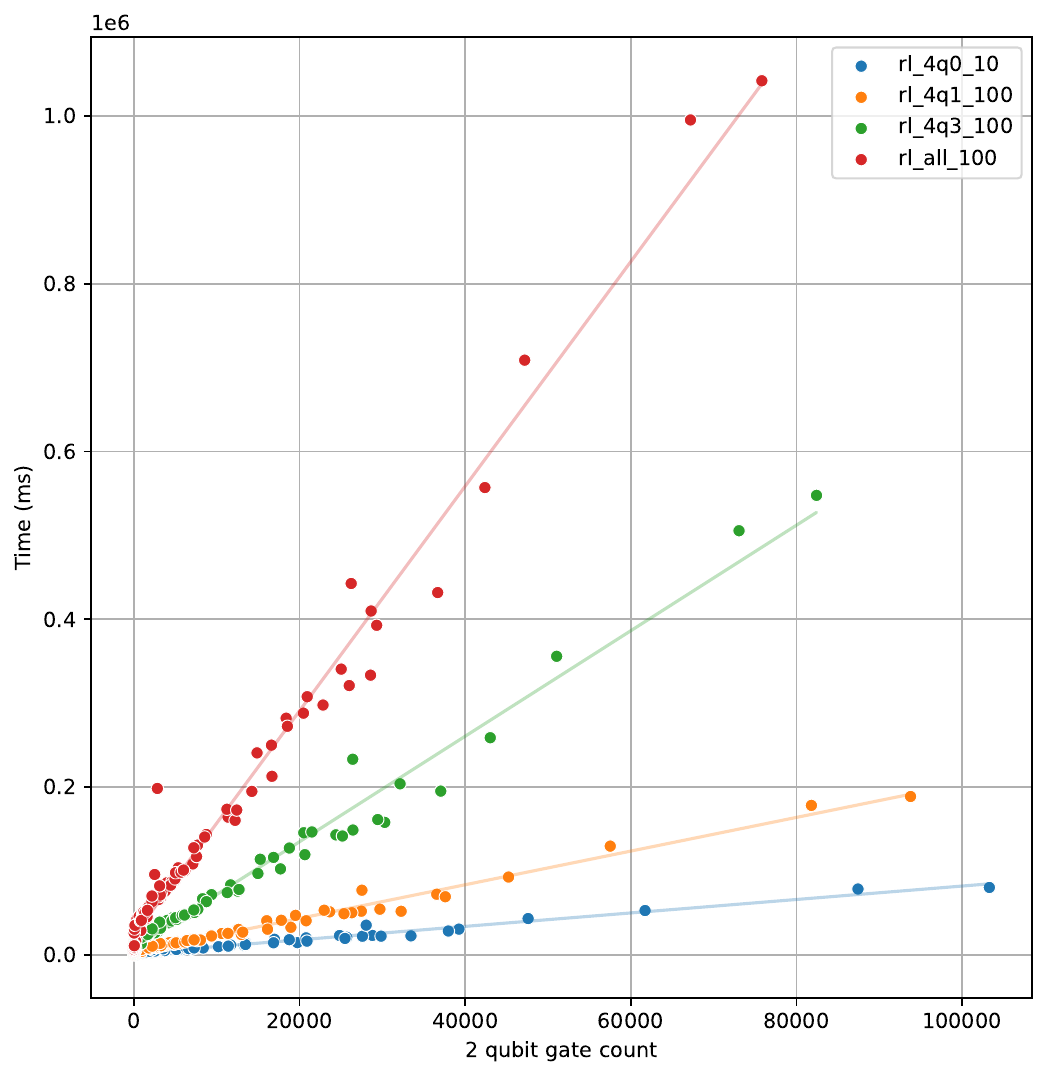}
    \caption{Total optimization runtime in milliseconds vs the number of two-qubit gates of the output circuit for the four different workflows. The time scales linearly with the number of gates in all cases, with more steep slopes for the more computationally intensive passes. }
    \label{fig:count_vs_time}
\end{figure}

Finally, we measure the runtime as a function of the number of two-qubit gates in Figure \ref{fig:count_vs_time}. This includes the time to do the collection, model selection and synthesis. We validate that the runtime scales linearly with the number of output two-qubit gates, with different slopes for the four workflows. As a reference, for the lightest optimization workflow, the optimization took less than 2 minutes for the largest circuits in Benchpress with over 1M+ two-qubit gates.

\begin{table*}[tp]
    \centering
    \begin{subtable}{\textwidth}

    \begin{tabular}{p{5.6cm}|p{2.3cm}p{2.3cm}p{2.3cm}p{2.3cm}}
    \toprule
    benchmark & \texttt{rl\_4q0\_10} & \texttt{rl\_4q1\_100} & \texttt{rl\_4q3\_100} & \texttt{rl\_all\_100} \\
    \midrule
    QASMBench - Small $(n \leq 10)$ & $0.92 \timesdiv 1.22$ & $0.89 \timesdiv 1.31$ & $0.86 \timesdiv 1.38$ & $0.83 \timesdiv 1.47$ \\
    QASMBench - Medium ($11\leq n \leq 27$) & $0.90 \timesdiv 1.12$ & $0.87 \timesdiv 1.16$ & $0.84 \timesdiv 1.19$ & $0.83 \timesdiv 1.21$ \\
    QASMBench - Large ($28\leq n \leq 433$) & $0.93 \timesdiv 1.08$ & $0.92 \timesdiv 1.10$ & $0.90 \timesdiv 1.12$ & $0.88 \timesdiv 1.14$ \\
    Feynman & $0.87 \timesdiv 1.15$ & $0.79 \timesdiv 1.18$ & $0.76 \timesdiv 1.20$ & $0.74 \timesdiv 1.21$ \\
    HamLib Benchpress & $0.81 \timesdiv 1.21$ & $0.76 \timesdiv 1.26$ & $0.72 \timesdiv 1.28$ & $0.69 \timesdiv 1.31$ \\
    \bottomrule
    \end{tabular}
    
    \caption{Depth improvement}
    \label{tab:depth_results}
    \end{subtable}
    \begin{subtable}{\textwidth}

    \begin{tabular}{p{5.6cm}|p{2.3cm}p{2.3cm}p{2.3cm}p{2.3cm}}
    \toprule
    benchmark & \texttt{rl\_4q0\_10} & \texttt{rl\_4q1\_100} & \texttt{rl\_4q3\_100} & \texttt{rl\_all\_100} \\
    \midrule
    QASMBench - Small $(n \leq 10)$ & $0.93 \timesdiv 1.17$ & $0.90 \timesdiv 1.24$ & $0.88 \timesdiv 1.29$ & $0.85 \timesdiv 1.37$ \\
    QASMBench - Medium ($11\leq n \leq 27$) & $0.92 \timesdiv 1.09$ & $0.89 \timesdiv 1.13$ & $0.86 \timesdiv 1.17$ & $0.85 \timesdiv 1.18$ \\
    QASMBench - Large ($28\leq n \leq 433$) & $0.94 \timesdiv 1.07$ & $0.93 \timesdiv 1.08$ & $0.92 \timesdiv 1.10$ & $0.90 \timesdiv 1.11$ \\
    Feynman & $0.88 \timesdiv 1.13$ & $0.82 \timesdiv 1.15$ & $0.79 \timesdiv 1.17$ & $0.77 \timesdiv 1.18$ \\
    HamLib Benchpress & $0.85 \timesdiv 1.19$ & $0.82 \timesdiv 1.2$ & $0.79 \timesdiv 1.21$ & $0.77 \timesdiv 1.21$ \\
    \bottomrule
    \end{tabular}
    
    \caption{Two-qubit count improvement}
    \label{tab:count_results}
    \end{subtable}
    \caption{Average improvement for circuits in the Benchpress benchmark after applying the four different optimization workflows to circuits transpiled with QiskitSDK transpiler (level 2).}
    \label{tab:benchpress_results}
\end{table*}

\tocless\section{CONCLUSIONS}\label{section:conclusions}

In this work, we presented an RL-based synthesis method for Pauli Network blocks of up to six qubits and demonstrated its use within a larger collect-and-re-synthesize framework for circuit optimization. By allowing for both Clifford operations and single-qubit Pauli rotations of arbitrary angles, our approach naturally incorporates parametric rotations and captures optimizations not possible when doing Clifford-only re-synthesis. We showed that, for small blocks, our RL procedure synthesizes circuits substantially shorter than existing heuristic methods, often reducing two-qubit gates and circuit layers by factors of two or more while still respecting hardware connectivity constraints. When integrated as a post-transpilation optimization pass within the QiskitSDK transpiler, the method delivered average improvements of 20\% or more in two-qubit gate count and depth across the Benchpress benchmarks, frequently exceeding 50\% in specific cases.

These results show the potential of RL-driven local circuit resynthesis for practical, large-scale quantum transpilation workloads. Going forward, a natural avenue for improvement is to develop more sophisticated collection strategies. Rather than naively selecting sub-circuits block by block, targeted collection of the parts of the circuit that contribute most to the global two-qubit depth (e.g., those on the “critical path”) could yield comparable or better optimizations with fewer sub-circuit extractions. Additionally, the use of more advanced commutation analyses within the collection process could identify larger or more relevant sub-circuits for which Pauli rotations may be reordered advantageously.

Another promising direction is to extend this framework to dynamic circuits featuring measurements and resets. By incorporating non-unitary operations into the RL-based synthesis routine, one could optimize mid-circuit measurements, conditional operations, and state reinitializations, ultimately broadening the applicability of RL-based methods for quantum systems and algorithms that leverage dynamic circuit capabilities. Through these enhancements and extensions, we expect RL approaches to play an increasingly central role in bridging high-level quantum algorithms with efficient, device-tailored circuit implementations.

\vspace{1.5\baselineskip}
\tocless{\section*{Author contributions}}

Conceptualization: D. K., A. D., S. M., D. W. Technical implementation: D. K., A. D., S. M., V. V. Training of RL algorithms: A. D., D. K. Execution of benchmarks: A. D., D. K. All authors contributed to the manuscript writing.

\tocless\acknowledgements
 The authors acknowledge insightful discussions and feedback received from Patrick Rall, Will Kirby, Arkopal Dutt and Elisa Baümer. The authors also acknowledge the IBM Research Cognitive Computing Cluster service for providing resources that have contributed to the research results reported within this paper.

\bibliography{biblio}


\clearpage
\setcounter{section}{0}  
\setcounter{figure}{0}   
\renewcommand{\thefigure}{S\arabic{figure}}

\onecolumngrid
\setcounter{page}{1}

\tocless{\section*{Appendix}}

\tocless{\subsection{Optimization Pass Results on Benchpress}}

\begin{figure}[h!]
    \centering
    \begin{subfigure}{\textwidth}
        \centering
        \includegraphics[width=\textwidth]{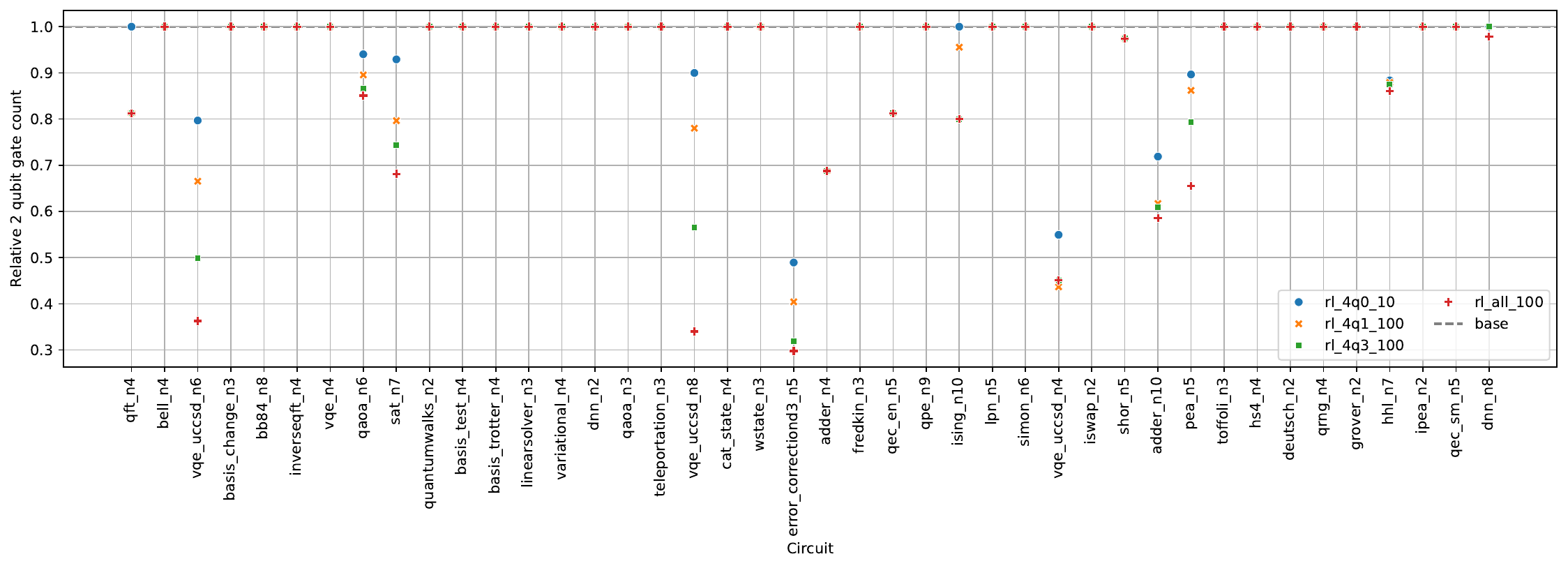}
        \caption{QASMBench small ($n\leq 10$)}
        \label{fig:qasmbench_small_count}
    \end{subfigure}
\hfill
    \begin{subfigure}{\textwidth}
        \centering
        \includegraphics[width=\textwidth]{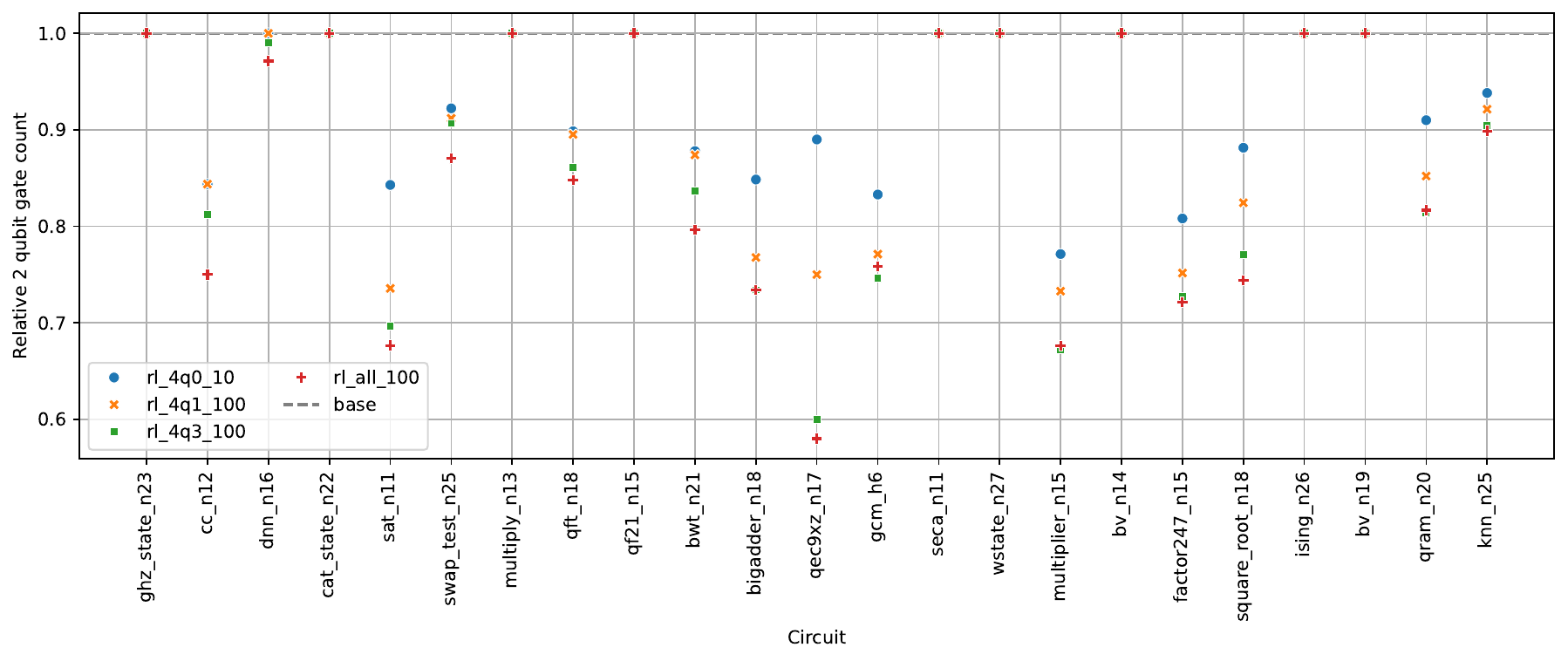}
        \caption{QASMBench medium ($11\leq n \leq 27$)}
        \label{fig:qasmbench_med_count}
    \end{subfigure}
\hfill
    \begin{subfigure}{\textwidth}
        \centering
        \includegraphics[width=\textwidth]{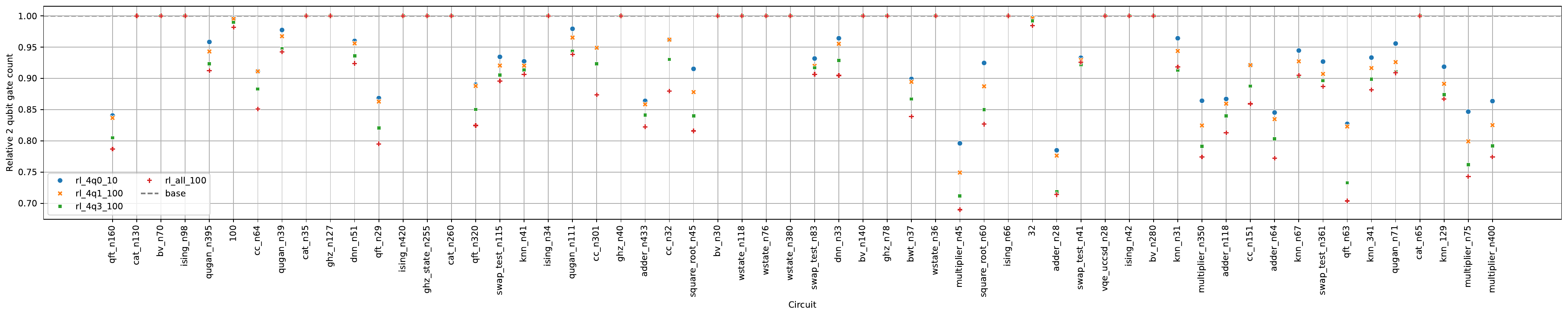}
        \caption{QASMBench large ($28\leq n\leq 433$)}
        \label{fig:qasmbench_large_count}
    \end{subfigure}
    \caption{Count results of optimization pass on QASMBench circuits}
    \label{fig:qasmbench_counts}
\end{figure}

\begin{figure}[h!]
    \centering
    \begin{subfigure}{\textwidth}
        \centering
        \includegraphics[width=\textwidth]{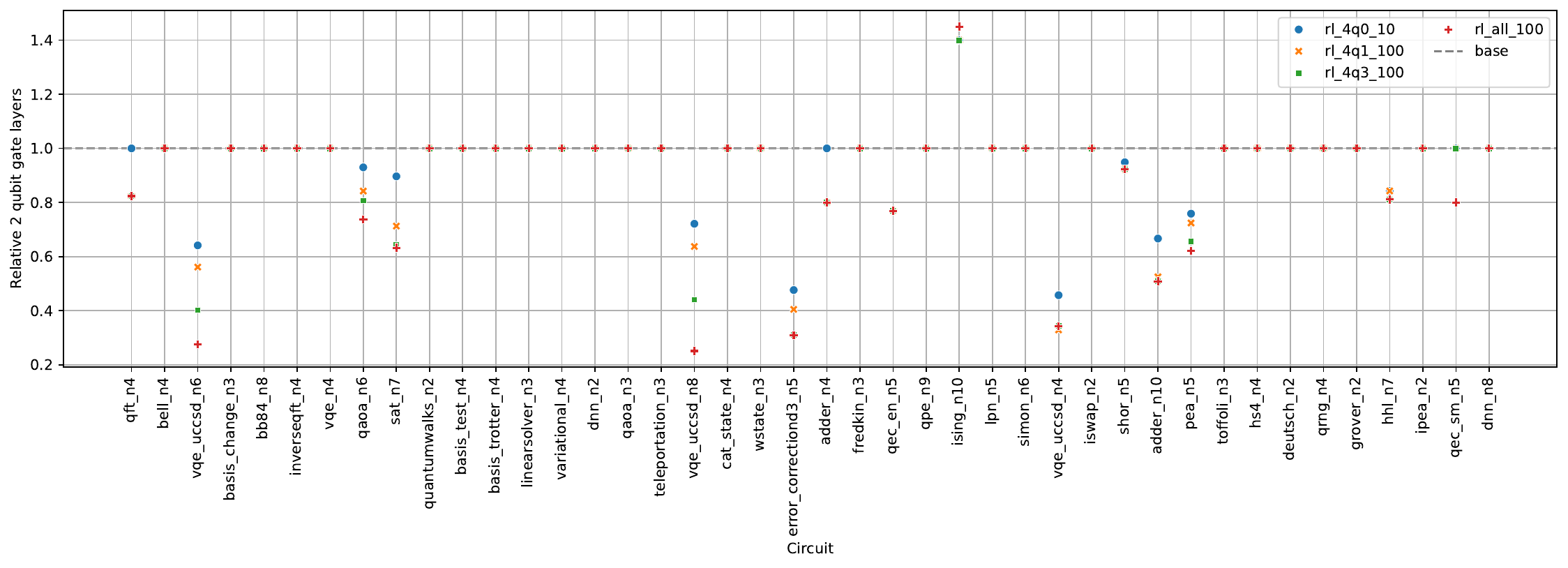}
        \caption{QASMBench small ($n\leq 10$)}
        \label{fig:qasmbench_small_depth}
    \end{subfigure}
\hfill
    \begin{subfigure}{\textwidth}
        \centering
        \includegraphics[width=\textwidth]{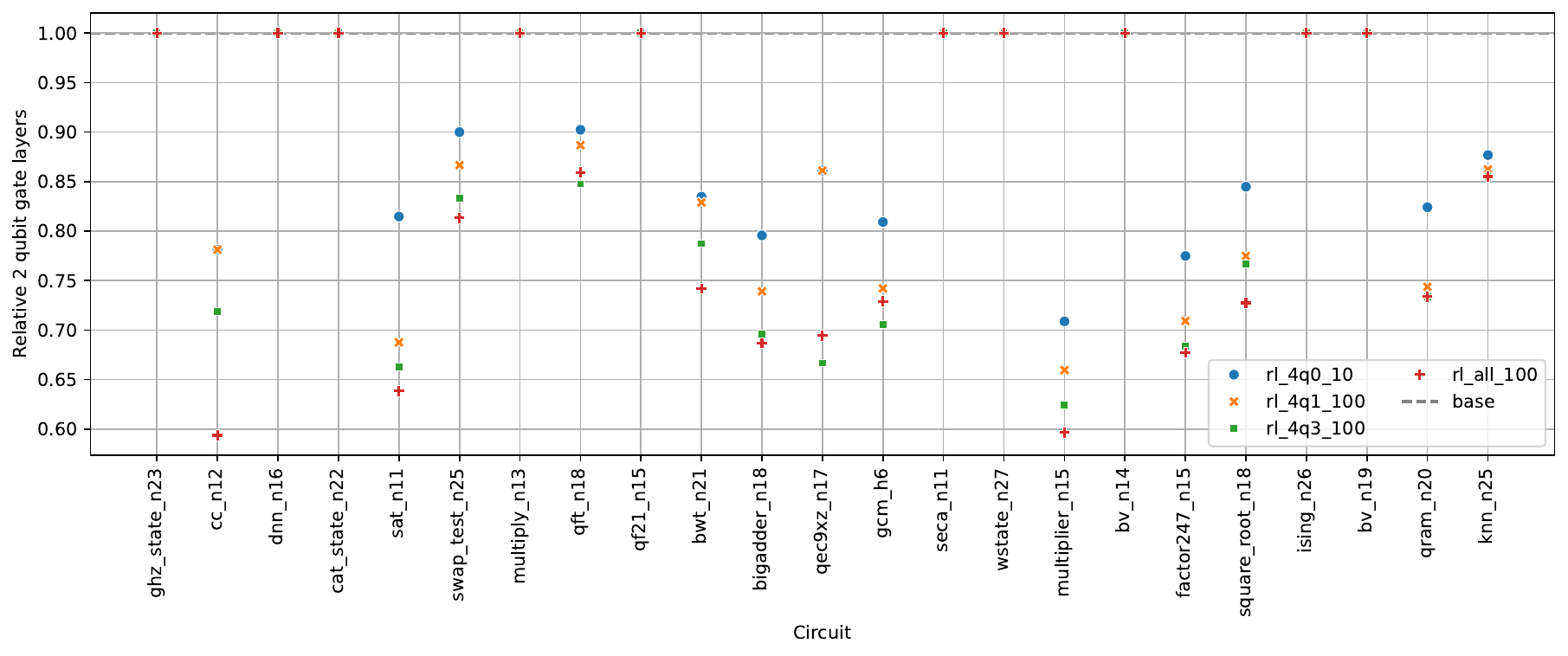}
        \caption{QASMBench medium ($11\leq n \leq 27$)}
        \label{fig:qasmbench_med_depth}
    \end{subfigure}
\hfill
    \begin{subfigure}{\textwidth}
        \centering
        \includegraphics[width=\textwidth]{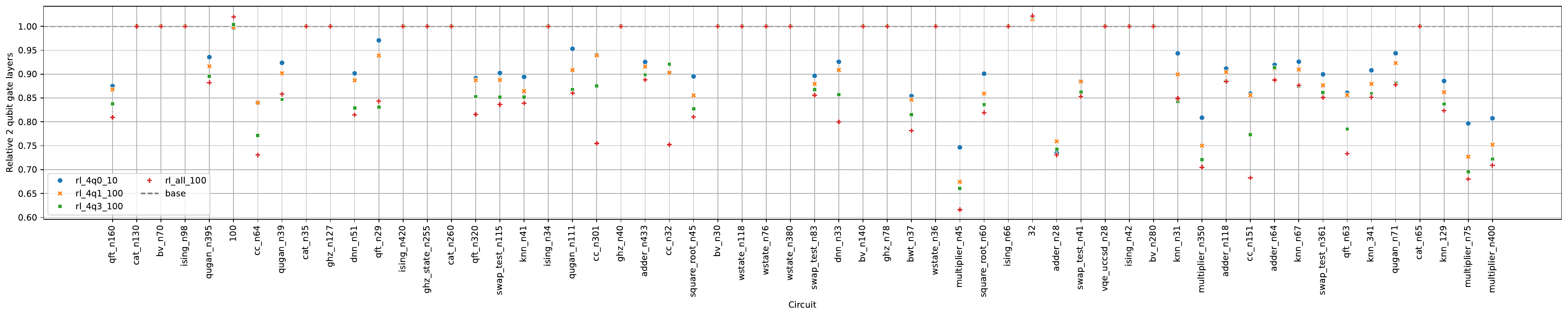}
        \caption{QASMBench large ($28\leq n\leq 433$)}
        \label{fig:qasmbench_large_depth}
    \end{subfigure}
    \caption{Depth results of optimization pass on QASMBench circuits}
    \label{fig:qasmbench_depths}
\end{figure}

\begin{figure}[h!]
    \centering
    \includegraphics[width=\textwidth]{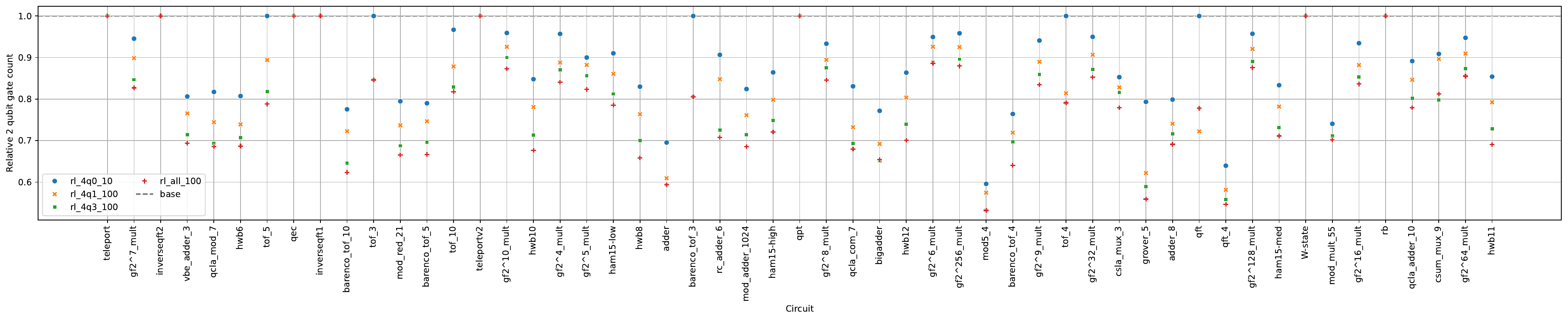}
    \caption{Count results of optimization pass on Feynman circuits}
    \label{fig:feynman_counts}
\end{figure}

\begin{figure}[h!]
    \centering
    \includegraphics[width=\textwidth]{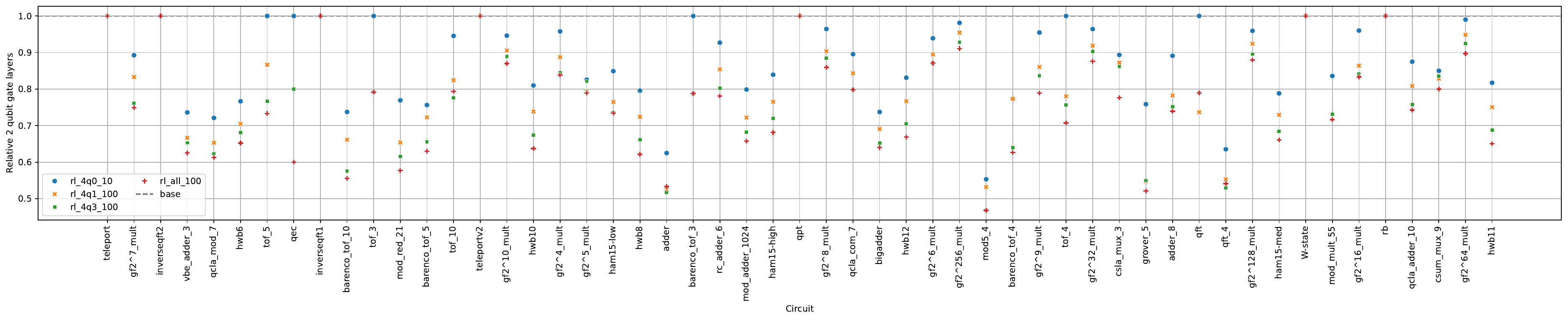}
    \caption{Depth results of optimization pass on Feynman circuits}
    \label{fig:feynman_depths}
\end{figure}

\begin{figure}[h!]
    \centering
    \includegraphics[width=\textwidth]{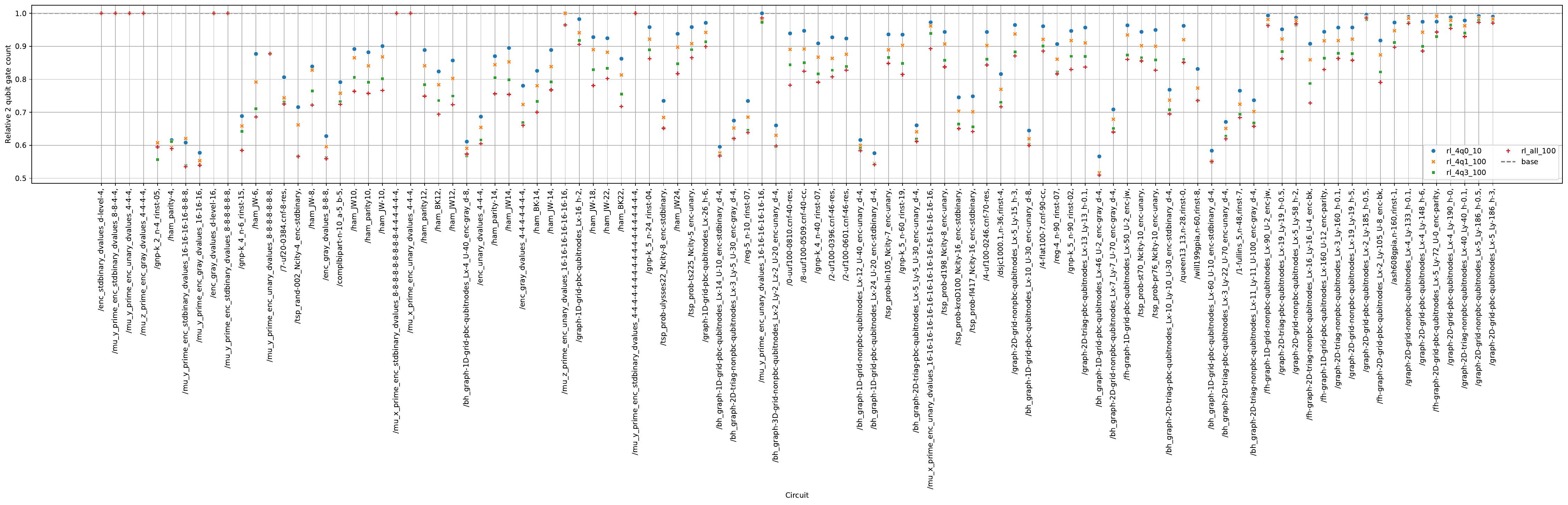}
    \caption{Count results of optimization pass on HamLib circuits}
    \label{fig:hamlib_counts}
\end{figure}

\begin{figure}[h!]
    \centering
    \includegraphics[width=\textwidth]{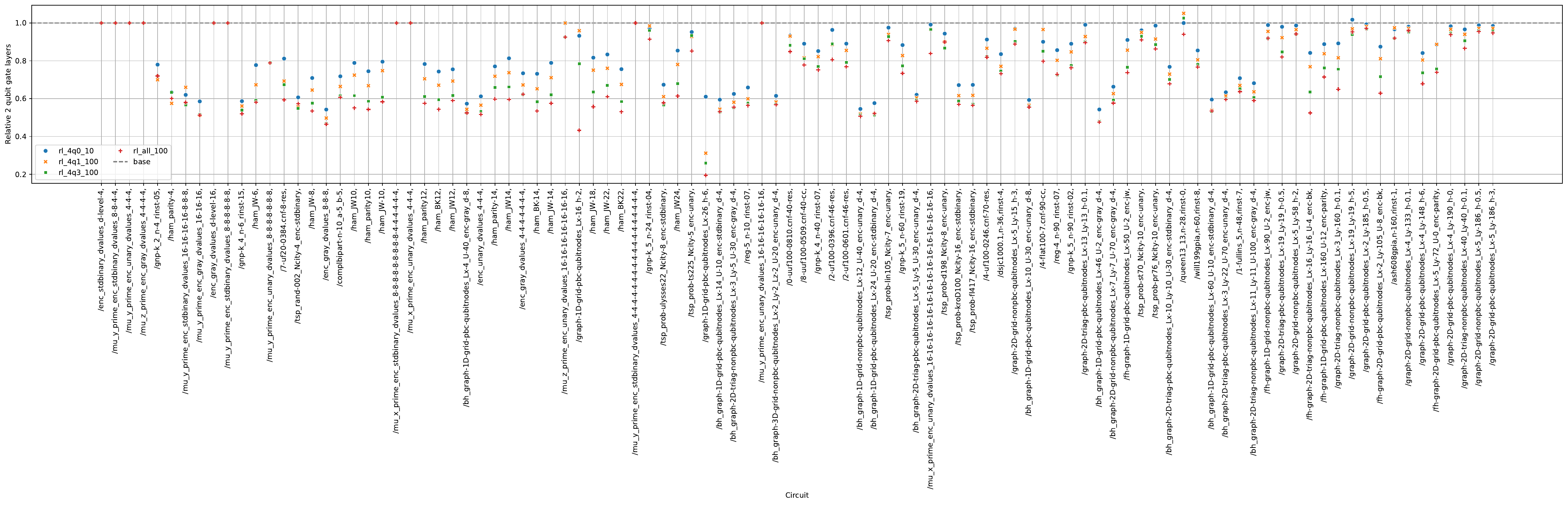}
    \caption{Depth results of optimization pass on HamLib circuits}
    \label{fig:hamlib_depths}
\end{figure}

\clearpage
\tocless{\subsection{Synthesis Benchmarks}}

\begin{table*}[h!]
    \centering
    \begin{subtable}[h!]{\textwidth}
    \begin{tabular}{p{2cm}|p{2.5cm}p{2.5cm}p{2.5cm}p{2.5cm}}
    \toprule
     \texttt{Coupling map} & \texttt{Qiskit\_Default} & \texttt{Qiskit\_Rustiq} & \texttt{RL\_1} & \texttt{RL\_10} \\
    \midrule
    4qL & 6.34 & 9.02 & 2.76 & 2.58 \\
    4qT & 7.27 & 6.86 & 2.68 & 2.48 \\
    5qL & 10.50 & 16.27 & 5.95 & 5.50 \\
    5qT & 11.27 & 12.88 & 5.43 & 5.02 \\
    6qL & 15.32 & 25.15 & 10.32 & 9.56 \\
    6qT & 15.88 & 21.18 & 9.11 & 8.40 \\
    6qY & 16.30 & 20.10 & 9.06 & 8.33 \\
    \bottomrule
    \end{tabular}
    \caption{Number of 2-qubit gates per Pauli}
    \end{subtable}
    \begin{subtable}[h!]{\textwidth}
    \begin{tabular}{p{2cm}|p{2.5cm}p{2.5cm}p{2.5cm}p{2.5cm}}
    \toprule
     \texttt{Coupling map} & \texttt{Qiskit\_Default} & \texttt{Qiskit\_Rustiq} & \texttt{RL\_1} & \texttt{RL\_10} \\
    \midrule
    4qL & 5.91 & 8.36 & 2.31 & 2.12 \\
    4qT & 7.27 & 6.86 & 2.64 & 2.45 \\
    5qL & 9.15 & 13.57 & 4.45 & 4.01 \\
    5qT & 10.51 & 11.70 & 4.53 & 4.14 \\
    6qL & 12.71 & 19.13 & 6.86 & 6.16 \\
    6qT & 13.99 & 17.26 & 6.50 & 5.84 \\
    6qY & 14.41 & 16.82 & 6.91 & 6.23 \\
    \bottomrule
    \end{tabular}
    \caption{Number of 2-qubit layers per Pauli}
    \end{subtable}

\begin{subtable}[h!]{\textwidth}
    \begin{tabular}{p{2cm}|p{2.5cm}p{2.5cm}p{2.5cm}p{2.5cm}}
    \toprule
     \texttt{Coupling map} & \texttt{Qiskit\_Default} & \texttt{Qiskit\_Rustiq} & \texttt{RL\_1} & \texttt{RL\_10} \\
    \midrule
    4qL & 2.57 & 2.35 & 0.40 & 1.79 \\
    4qT & 2.58 & 2.31 & 0.35 & 1.41 \\
    5qL & 3.16 & 2.90 & 0.67 & 3.14 \\
    5qT & 4.11 & 2.70 & 0.61 & 2.91 \\
    6qL & 4.17 & 3.73 & 1.19 & 6.15 \\
    6qT & 3.95 & 3.45 & 1.08 & 5.86 \\
    6qY & 3.97 & 3.49 & 1.06 & 5.16 \\
    \bottomrule
    \end{tabular}
    \caption{Time taken (in milliseconds) per Pauli}
    \end{subtable}
    
    \caption{Synthesis+Transpilation workflows benchmarked on randomly generated Pauli Networks containing between 0 and 100 rotations and as many  qubits as the respective coupling map.}
\end{table*}

\end{document}